\documentclass[12pt]{article}

\usepackage[textheight=240mm,textwidth=169.69mm]{geometry}
\usepackage{graphicx} % Required for inserting images
\usepackage{natbib}
\usepackage{amsmath,amssymb,amsfonts,amsthm}
\usepackage{xcolor}
\usepackage{multirow}
\usepackage{makecell}
\usepackage{siunitx}
\usepackage{booktabs}
\usepackage{hyperref}
\hypersetup{
    colorlinks=true,
    linkcolor=blue,
    filecolor=blue,      
    urlcolor=blue,
    citecolor=blue
    }
\usepackage{anyfontsize}

\usepackage{newtxtext}
\usepackage{newtxmath}

\newcommand{\RomanNumeralCaps}[1]
\title{Fast and slow surfactants in turbulence-driven bubble breakup}
\author{Zhan Wu$^1$\footnote{Author contributions: Z.W. and T.A. contributed equally to this work.}~,
  Tristan Aurégan$^1$\footnotemark[1] ~, and Luc Deike$^{1,2}$}

\date{\small $^{1}$Department of Mechanical and Aerospace Engineering, Princeton University, Princeton, NJ, USA\\
$^{2}$ High Meadows Environmental Institute, Princeton University, Princeton, NJ, USA}

\begin{document}
\maketitle

\begin{abstract}

When a large air cavity breaks in a turbulent flow, it goes through very large deformations and cascading events of new interface formation, including elongated filaments and bubbles over a wide range of scales, with their rate of formation controlled by turbulence and capillary processes. We experimentally investigate the effects of surfactants and salt on the fragmentation of a large air cavity, and observe a five fold increase of the number of bubbles being produced in some cases. We characterize the bubble size distribution resulting from successive break-ups in such transient configuration. For bubbles larger than the Hinze scale $d_H$ (defined as the balance between surface tension and turbulence stresses), we observe that bubble size distributions remain unchanged for all solutions tested. For bubbles below $d_H$, however, we observe an increase of the number of bubbles produced and an associated steepening of the bubble size distribution upon the addition of surfactant or salt. This later effect is only visible for surfactants with an adsorption timescale that is fast enough compared to the rate at which new interfaces are being generated by turbulence.
\end{abstract}

%\begin{keywords}
%Bubbles, fragmentation, surfactants, adsorption
%\end{keywords}

%{\bf MSC Codes }  {\it(Optional)} Please enter your MSC Codes here

\section{Introduction}
Gas bubbles dispersed in turbulent flows play a crucial role in mediating the transfer of mass and energy across {interfaces \citep{Mathai2020}. At the ocean surface, the bursting of bubbles produces sea spray aerosols \citep{Deike2022, Deike2022agu,Cochran2017}, influencing atmospheric and climate processes, while bubbles entrained underwater break \citep{Deike2016,Deane2002} and influence gas exchange \citep{deike2018gas,Deike2022,deike2025}. Underwater bubbles transport facilitates oil and gas migration from deep wells, while bubble breakup enhances gas dissolution through increased interfacial area and enhanced mixing in bubble column reactors \citep{Galinat2005,Risso2018}. Bubbles also play an important role in wastewater treatment, energy systems, advanced manufacturing, food production and agriculture \citep{Garbin2025bubbly}.

In practical configurations, air-water interfaces are almost always contaminated by surface active agents (surfactants) related to biological or anthropogenic activities. The presence of surfactant modulates surface waves dynamics \citep{Erinin2023}, bubble rise velocity \citep{takemura1999rising,tagawa2014surfactant,farsoiya2024}, bubble bursting \citep{eshimaauregan2025}, with surface tension gradients leading to Marangoni flows, as well as tip streaming leveraged in microfluidics devices to produce droplets \citep{eggleton2001tip,rubio2023influence}.
Another practical example of surfactant effects include oil-contaminated surfaces, where surfactants are used in mitigation strategies but also lead to chemical transport by aerosolization \citep{Sampath2019, Feng2023}. Beyond Marangoni effects, variations in surface tension and sorption kinetics also play important roles \citep{fernandez2025influence}, and comparison between the time scale of adsorption/desorption of surfactant molecules with the characteristic time scale of the flow of interest becomes necessary. A comprehensive understanding of surfactant-mediated bubble behavior is essential for both mitigating environmental impacts of marine pollution and optimizing industrial processes involving gas–liquid interactions.

%In realistic conditions, bubble dynamics are strongly affected by the presence of surfactants. For example, surfactants and dispersants promote the adsorption of long-chain n-alkanes onto bubble or droplet surfaces through thermodynamically favorable interactions, thereby reducing oil plume formation during bubble bursting \citep{LiyanaArachchi2014}. The Marangoni effect can further modulate the bursting process and control the size of ejected droplets (Eshima 2025), while variations in surface tension and sorption kinetics also play important roles (Martinez 2025). 

Here, we consider the effect of surfactant on bubble break-up in a transient configuration, where a large air cavity breaks into small bubbles through successive fragmentation events. The configuration is largely inspired by air entrainment and bubble fragmentation under breaking waves. Bubble fragmentation under breaking waves or in turbulent flows has been widely investigated through both experiments \citep{Deane2002,Ruth2022,ni2024deformation} and numerical approaches \citep{Deike2016,mostert2022,Riviere2021,Riviere2022}. Some numerical studies have probed the effect of surfactants in two-phase turbulence \citep{soligo2019breakage,cannon2024morphology}, and a few experiments investigated the role of salt in bubbly flows \citep{blaauw2025salts}, but to the best of our knowledge, no systematic studies on the role of surfactants on the distribution of bubbles in turbulence exist.

\begin{figure}
    \centering
    \includegraphics[width=0.9\linewidth]{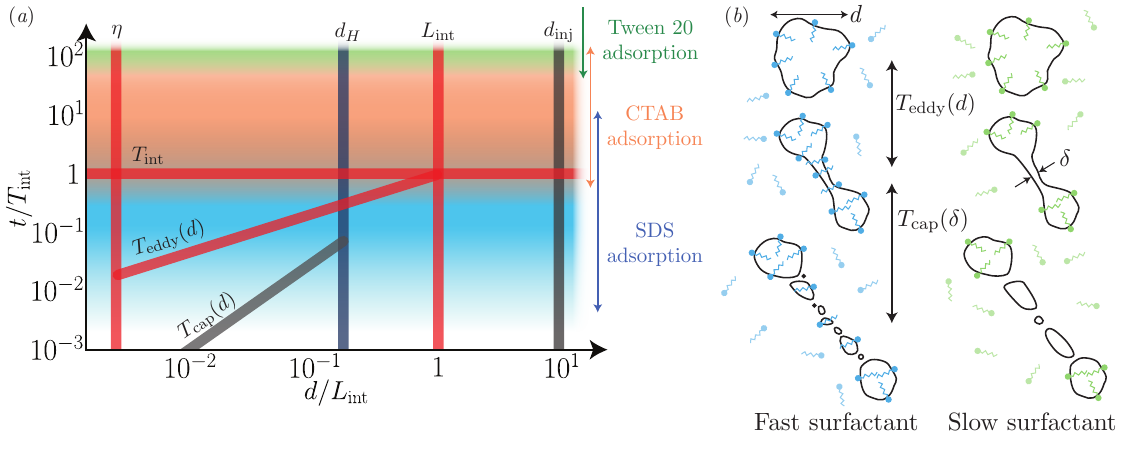}
    \caption{(a) Comparisons of the timescales of turbulence, capillary instability and adsorption of the three surfactants used in this study. The length scales are normalized by the integral length scale of turbulence $L_{\rm int}$ and the time scales by the integral time $T_{\rm int}$. $\eta$ is the Kolmogorov scale and $d_{\rm inj}$ is the equivalent diameter of the volume of air initially injected. (b) Schematic of the adsorption dynamics of a fast or slow surfactant during the break-up process. An bubble of  size $d$ intially covered with surfactants is stretched forming filaments, over a typical time $T_{\rm eddy}(d)$. The air filament of radius $\delta$ fragments in to small bubbles in a time of the order $T_{\rm cap}(\delta)$.}
    \label{fig:1}
\end{figure}

Bubble breakup in turbulence is understood by considering the competition between inertial stresses from turbulence that promote fragmentation and surface tension forces that resist it. The Weber number, \( \text{We} = \rho U^2 d / \gamma_s \), which compares these two effects, is commonly used to characterize bubble breakup conditions. Here, \( d \) is the bubble’s volume-equivalent diameter, \( U \) a characteristic velocity defined as the average velocity difference at the bubble-diameter scale, \( \rho \) the liquid density, and \( \gamma_s \) the surface tension. For bubbles within the inertial subrange, the square of the characteristic velocity at the bubble scale is $2 \epsilon^{2/3} d^{2/3}$, leading to $\text{We} = 2\rho \epsilon^{2/3} d^{5/3}/ {\gamma_s}$ \citep{Riviere2021}.
A critical Weber number $We_c$ provides the balance between inertial and surface tension forces, serves as a loose bubble breakup criterion and is of order unity \citep{MartinezBazan1999, Vejrazka2018,Riviere2021}. The associated length scale is the Hinze scale, $d_H$ \citep{Hinze1955}:
\begin{equation}\label{eq:dh}
d_H = \left(\frac{We_c}{2}\right)^{3/5}
\left(\frac{\gamma_s}{\rho}\right)^{3/5}
\epsilon^{-2/5},
\end{equation}
which separates the super-Hinze regime ($d>d_H$) in which bubbles will break due to turbulence, from the sub-Hinze regime ($d<d_H$). The critical Weber number or Hinze scale is a soft threshold due to the inherent stochasticity of turbulent flows. The super-Hinze bubble size distribution in turbulence or under breaking waves has been described by \( \mathcal{N}(d) \propto d^{-10/3} \), explained by a turbulent break-up cascade, controlled by the eddy turn over time at the scale of the bubble \( T_{\text{eddy}}(d) = \epsilon^{-1/3} d^{2/3} \) \citep{Garrett2000}. In contrast, the sub-Hinze bubble size distribution exhibits a gentler slope, following $\mathcal{N}(d)\propto d^{-3/2}$, which can be explained by a capillary break-up process related to large bubbles experiencing large deformation and the formation of filaments which will then rupture in a time scale controlled by capillarity \( T_{\text{cap}}(d) = (2\sqrt{3})^{-1} \left(\rho/\gamma_s\right)^{1/2} d^{3/2} \) \citep{Riviere2022,Ruth2022}. Both regimes have been extensively documented through laboratory experiments \citep{Ruth2022,Deane2002, qi2022fragmentation} and direct numerical simulations \citep{mostert2022,Riviere2022, Crialesi2023, King_2023}. The formation of such broad size distribution from an initial air cavity occurs through a sequence of events controlled by the turbulent and capillary time scales, with new interfaces formed along the way.

Figure~\ref{fig:1} (a) illustrates the time scales of bubble break-up in turbulence as a function of bubble size $d$, using the integral length scale and turbulent dissipation rate estimated in our laboratory experiments. The lengths are normalized by the integral length scale $L_{\rm int}$, corresponding to the length scale of the larges eddies ($\approx 2.5$~cm) and the times are normalized by the associated integral timescale ($T_{\rm int} \approx 100$~ms).

%Discuss general motivation, bubbles in environment and engineering, turbulent flows. Discuss presence of contamination in most practical problems. How much does it matter? Discuss specific examples of effect of surfactant.

%The physics of bubble breaking in turbulence.
%The Hinze scale, $d_H$ is defined as
%\begin{equation}
%d_H = \left(\frac{We_c}{2}\right)^{3/5}
%\left(\frac{\gamma_s}{\rho}\right)^{3/5}
%\epsilon^{-2/5},
%\end{equation}
%where $We_c = 1$ following Riviere (2022). 

%Discuss briefly the two power law regimes and their physics, sub and super Hinze scale regimes.

When using surfactants, we now have to consider their relevant time scales and how they compare to capillary and turbulence times which will control break-up and deformation. Marangoni stresses related to gradients in surfactant concentration along the interface will occur on capillary time scales. However, the presence and amount of surfactant on a newly formed interface within a sequence of break-up events will be related to the surfactant adsorption time $\tau_{\rm ads}$, which controls how fast surfactant in the bulk will make it to the interface \citep{manikantan2020surfactant}. As we will demonstrate experimentally, surfactants will influence the multi-scale break-up processes in different ways depending on the characteristic adsorption time of the surfactant, illustrated for Sodium Dodecyl Sulfate (SDS, fast adsorption), Tween~20 (slow adsorption) and Cetrimonium bromide (CTAB, intermediate adsorption timescale) in figure \ref{fig:1} (a). For SDS, which has a short adsorption time of the order of  miliseconds to seconds depending on the concentration \citep{chang1995adsorption,Fainerman2010}, the sorption dynamics will be occurring over a time scale comparable to $T_{\rm eddy}$, allowing SDS molecules to adsorb onto the surface of newly formed bubbles or filaments and influence the breakup process. However, the capillary instability leading to the final break-up will occur on time scales faster than the adsorption. In contrast for Tween~20, which has a long adsorption time of the order of tens of seconds \citep{bkak2016interfacial,Qazi2020}, $T_{\rm eddy}$ and $T_{\rm cap}$ are generally shorter than its adsorption time, leaving insufficient time for adsorption during a single set of breakup event. With an intermediate adsorption timescale of order one second \citep{Qazi2020}, CTAB lies between SDS and Tween~20. Its adsorption time is comparable to the integral timescale, $T_{\mathrm{int}}$, allowing adsorption during the breakup process. Consequently, CTAB behaves similarly to Tween~20 at low concentrations and transitions towards SDS-like behaviour as the concentration increases. This observation (illustrated in figure \ref{fig:1}b) will guide the interpretation of our experiments varying surfactant concentration leading to comparable changes in the static surface tension. We note however that the process described here: formation of a dumbbell shape, then ligament rupture; is an idealization of the many processes resulting in sub-Hinze bubble production. Other processes such as direct tearing from a large bubble for instance can also occur but they also involve the formation of a large surface area elongated structure, before capillary rupture.

The paper is organized as follows. Section 2 presents the experimental techniques while Section 3 presents the resulting bubble size distributions, the conclusions are discussed in Section 4.

\section{Experimental methods}

\begin{figure}
    \centering
    \includegraphics[width=\linewidth]{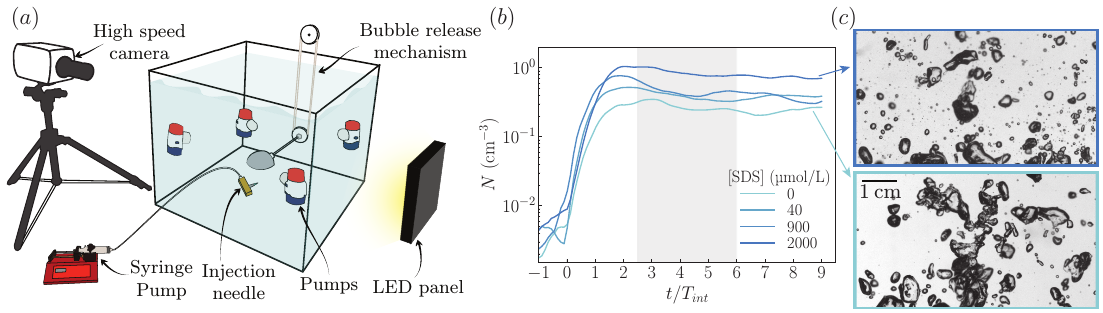}
    \caption{(a) Schematic of the experimental setup: air is injected into an upside down cup that is flipped at the beginning of the experiment, and a large air cavity is released into the turbulent flow. (b) Temporal evolution of the total bubble count $N$ for different SDS concentrations in deionized water. (c) Example of images comparing clean (light blue, bottom), and contaminated cases (dark blue, top, 2000~\textmu mol/L of SDS). The shaded area ($2.5T_{\rm int}\sim 6T_{\rm int}$) represents the time range used to compute distributions in the following.}
    \label{fig:2}
\end{figure}

\subsection{Experimental Set-up}
Figure \ref{fig:2} (a) shows the schematic diagram for the experimental setup, similar to \cite{Ruth2022}. It consists of a polycarbonate tank filled with about 150 L of solution, and a hollow hemispherical metal cup is positioned underwater at the tank center, equidistant from the four corners of the tank where four pumps (Rule 20 DA 800 GPH Bilge Pump) are mounted. The cup initially faces downwards and is filled with a constant volume $V_0 = 20$~mL of air. At the beginning of an experiment, the cup is rotated to face upwards while a fast camera (Phantom VEO4K-990-L, 500 fps) is simultaneously triggered. 
%The cup is initially facing downwards, allowing air bubbles injected from a fixed needle at the tank bottom to accumulate and coalesce into a single air cavity inside it. The volume of the air cavity $V_0$ = 20 mL is controlled by a syringe pump and kept constant throughout the study. The cup is mounted on a metal shaft connected to a rotating gear and belt system. A magnet attached to the belt passes a Hall effect sensor when the cup is rotated a quarter of a revolution, triggering a high-speed camera (Phantom VEO4K-990-L) for synchronized recording. Upon rapid half-revolution inversion of the cup, the single air cavity is released into the turbulent flow 

The air cavity is released into the turbulent flow and experiences a sequence of breakups which results in a broad distribution of bubble sizes. Such experimental configuration is designed to characterize the transient fragmentation process of a large air cavity into a distribution of bubbles of various sizes. The high-speed camera is coupled with two different lenses to improve the resolution using multiple views: a large scale view (Nikon Micro 200mm leading to 35~\textmu m/pixel) and a zoomed-in view (K2 DistaMax Infinity Photo-Optical Co. lens leading to 9~\textmu m/pixel). The field of view is illuminated by a coaxial LED panel from the opposite side. %The frame rate of the high-speed camera is 500 Hz.
To ensure statistical convergence of the size distribution, each experimental condition is repeated 15 times using the 200mm lens and 25 times using the K2 lens. The depth of field $\delta(d)$ was calibrated by translating a target containing 14 chrome-sputtered dots with diameters ranging from 30 µm to 3000 µm along the optical axis, similarly to \cite{erinin2023method}. The large scale field of view is processed with a two step algorithm: thresholding for larger bubbles then edge detection on an image where large bubbles have been removed to detect the smaller ones. The small scale field of view is processed using thresholding followed by a filter on the sharpness of the detected objects, to only keep well focused bubbles.
We therefore obtain the bubble size distribution $\mathcal{N}(d)$, defined as the number of bubbles per unit bin size $[d, d+\Delta d]$, normalized by the cross sectional area $A$ of the camera and the depth of field at that given size bin $\delta(d)$, and averaged over all realizations. The distributions from the two camera views are combined into a single distribution (see \cite{Mazzatenta2025}). 
The uncertainties in the distribution is estimated by a Monte Carlo approach in which data are randomly sampled from the repeated experiments using bootstrapping (100 realizations). %The mean and uncertainty of the bubble size distribution $\mathcal{N}(d)$ are shown as solid lines and shaded regions in figure~\ref{fig:3}(a–c) respectively. Furthermore, we integrated $\mathcal{N}(d)$ over $150~\mu\mathrm{m} \le d \le d_{H}$ to get the sub-Hinze bubble counts and over $d_{H} \le d \le 20~\mathrm{mm}$ to get the super-Hinze bubble counts. The uncertainty of the normalized bubble counts, $N/N_0$, where $N_0$ is the reference bubble counts in clean DI water, was computed as \( (\Delta({N}/{N_0}))^2=(N/N_0)^2[(\Delta N/N)^2+(\Delta N_0/N_0)^2] \).

%\begin{equation}
%\mathcal{N}(d)=\frac{\mathrm{N}(d)}{\Delta d\cdot \delta(d)\cdot A\cdot \Delta t\cdot N_{\mathrm{exp}}},
%\label{eq:1}
%\end{equation}
%where $\mathrm{N}(d)$ is the total number of bubbles of diameter within the bin $[d, d+\Delta d]$ detected across $N_{\mathrm{exp}}$ realizations during the time interval $\Delta t$, and $A$ is the cross-sectional area of the camera’s field of view.

\subsection{Surfactant and salt conditions}

\begin{table}
  \begin{center}
\def~{\hphantom{0}}
  \begin{tabular}{lcccccc}
      Surfactant &
      Sea Salt &
      Concentration (\% CMC) &
      $\gamma_s$ (mN/m) &
      $d_H$ (mm) &
      $d_{32}$ (mm) \\[3pt]

      SDS     &  No           & 0--80   & 70--37 & 2.1--3.1 & 5.2--6 \\
      SDS     & Yes & 0--1.2  & 65--39 & 2.1--2.9 & 6--7.3 \\
%      SDS     & tap water            & 0--2    & 72--55 & 2.6--3.1 & 6.1--6.8 \\
%      SDS     & No          & 0--80    & 72--42 & 2.3--3.1 & 5.7--7.5 \\
      Tween-20 & No            & 0--100  & 72--41 & 2.2--3.1 & 6.4--6.9 \\
      Tween-20 & Yes & 0--40  & 72--43 & 2.3--3.1 & 6.7--8.1 \\
      CTAB & No             & 0--90  & 71--38 & 2.1--3.1 & 5.4--6.8 \\

  \end{tabular}
  \caption{Summary of surfactant and solvent conditions tested in the study, with the corresponding evolution of the measured static surface tension $\gamma_s$, the Hinze scale $d_H$ computed using \eqref{eq:dh}, and the Sauter mean diameter of the measured distributions $d_{32}$. Concentrations are given as a fraction of the Critical Micelle Concentration (CMC): 8.2~mmol/L for SDS, and 59~\textmu mol/L for Tween 20 and 1~mmol/L for CTAB.}
  \label{table:1}
  \end{center}
\end{table}

There are five different sets of solutions in this study. We used three different surfactants: SDS, CTAB, and Tween 20 used as purchased and mixed with deionized water. We also repeated experiments with SDS and Tween 20, adding a mixture of salts representative of salts present in the ocean (ASTM D1141-98 Artificial Sea Salt, same as in \cite{Mazzatenta2025}). The case with SDS was repeated a second time to ensure robustness. For each experimental case, the solution’s surface tension isotherm was measured at the beginning and end of the experiment using a Langmuir trough (KSV NIMA 1003), and the static surface tension prior to surface compression was recorded as $\gamma_s$. Surfactants concentrations were chosen such that a given set of experiments covers the whole range from clean interfaces to fully contaminated (concentration close to the CMC or surface tension close to 40~mN/m).
%The lowest surface tension achieved across all experiments was recorded as $\gamma_{\mathrm{cmc}}$, which was obtained by adding 80$\%$ CMC of SDS into DI water. 
During all experiments, the tank was closed with an acrylic lid to minimize contamination from ambient air. Experimental conditions are summed up in table \ref{table:1}. %The choice of solution was made to explore the importance of salt, as well as surfactants with very different adsorption time (SDS: fast; Tween~20: slow; CTAB: intermediate) as illustrated in figure \ref{fig:1}. 

Figure~\ref{fig:2}(b) presents the temporal evolution of the bubble count, $N = \int \mathcal{N}(d)\mathrm{d}d$ %with $L=220$ mm—the distance between the outlets of two adjacent pumps
, for experiments conducted with varying concentrations of SDS in deionized water. 
Across all experimental conditions, $N$ exhibits a sharp increase immediately after the cup is flipped ($t=0$), followed by a quasi-steady plateau over the interval $2.5T_{\mathrm{int}} \le t \le 6T_{\mathrm{int}}$. %, where $T_{\mathrm{int}}$ denotes the integral timescale of the turbulent flow. 
The total number of recorded bubbles increases with the concentration of SDS by almost an order of magnitude.
Figure~\ref{fig:2}(c) compares images of the clean case (no surfactants, bottom) and a contaminated case ([SDS] = 2000 \textmu mol/L, top) at $t = 4T_{\mathrm{int}}$. The scale bar represents 1 cm. A substantial increase in the number of small bubbles is observed in the contaminated case, whereas the population of large bubbles remains nearly unchanged.
Similar observations are made when considering the other types of solutions in terms of time evolution (shown in Appendix). 
Bubble coalescence in the present experiment is expected to be negligible given the low volume fraction of air, as discussed in experiments \citep{Ruth2022} and in simulations \citep{Riviere2021,Riviere2022}. 

We note that the experiment is transient by design as a large cavity is released and breaks due to turbulence. In our experiment, we observe bubbles generated over a relatively short window after the release (about 1s, or $10 T_{\rm int}$). During that time, bubbles rise to the surface and break and we observe an initial increase in the observed bubbles which plateaus for some time, and would then decrease at much longer times if we were continuing to record as all bubbles get out of the experimental field of view. As we analyze the influence of surfactant on bubble break-up, we will compare the adsorption time with the time in between the break-ups in the sequence, as new area is constantly being created through successive breakups.

%In the case of DI water with Tween~20, the total number of bubbles is very similar across all contaminations, a feature we will discuss in details when considering the bubble size distribution.

%$N_0$ and $\gamma_0$ denote the reference bubble count and reference surface tension in clean DI water.

\begin{figure}
    \centering
    \includegraphics[width=\linewidth]{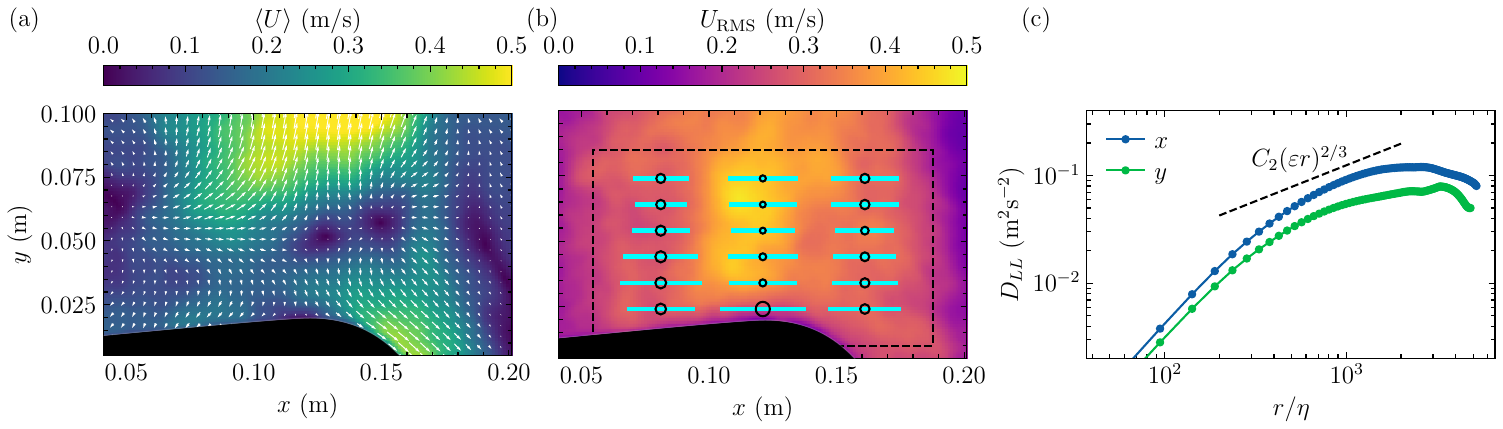}
    \caption{Properties of the turbulent flow field in the setup. Panel (a) shows th mean flow $\left< U \right>$ intensity and direction. Panel (b) shows the RMS velocity in color. The dashed rectangle represents the field of view, the cyan lines show local estimations of the integral length scale $L_{\rm int}$ and the circles show the local Hinze scale $d_H$. Panel (c) shows the second order longitudinal structure functions in both directions. The dashed line is the inertial range scaling, with the prefactor $C_2=2.0$.}
    \label{fig:piv}
\end{figure}

\subsection{Turbulent properties}

We measured the turbulent properties of the flow using Particle Image Velocimetry (PIV) in the central plane of the experiment, just over the cup with a field of view slightly larger than the one used in the experiment. The setup, processing and analyses are very similar to the ones described in \citet{Ruth2021}. We recorded the flow seeded with tracer particles for 2.5~s with a sampling rate of 1000 frames per second. The data is processed using a standard library for PIV analysis. The mean flow is similar to the one used in \citet{Ruth2022}: a small downwards flow just above the cup due to the jets of the pumps meeting in the center (Fig. \ref{fig:piv} (a)). 
We estimated the turbulent properties of the flow using the scaling argument in homogeneous and isotropic turbulence \citep{dejong_dissipation_2009}: $\varepsilon=AU_{\rm RMS}^3/L_{\rm int}$ with $U_{\rm RMS}$ the root-mean-square of the velocity fluctuations and $L_{\rm int}$ the integral length scale, using $A=0.7$ \citep{sreenivasan_update_1998}. We compute the integral length scale as the maximum of the integral of the autocorrelation of the velocity fluctuations, and average between both directions. Since there is some inhomogeneity in the flow structure, we computed the integral length scale and RMS velocity locally, giving us a local dissipation rate. The resulting integral length scale (cyan lines) and Hinze scale (black circles, using Eq. \eqref{eq:dh}) are plotted in Fig. \ref{fig:piv} (b) in several locations in space.
The integral length scale is $L_{\rm int} =2.5$~cm (10\%: 2.2, 90\%: 2.8), the root-mean-square velocity is $U_{\rm RMS}= 34$~cm/s (10\%: 28, 90\%: 42), giving a dissipation rate $\varepsilon=0.39$~m$^2$s$^{-3}$ (10\%: 0.22, 90\%: 0.75) where the value reported is the median of the spatial distribution and the values in parentheses are the quantiles, showing some inhomogeneity in our setup. The associated Kolmogorov scale is 40~\textmu m and the integral timescale is $T_{\rm int}=101$~ms.
In Fig. \ref{fig:piv} (c), we show the second order longitudinal structure function $D_{LL}(r)=\left< (\Delta u(r))^2\right>_{\mathbf{x}, t}$, in both $x$ and $y$ directions. In addition we plot the inertial range scaling, with the dissipation computed from the approach above, and the prefactor $C_2=2.0$ \citep{Pope2000}. Despite the limited inertial range in our setup, we see that both approaches are consistent.
Overall, the flow field in the setup presents strong turbulent fluctuations, required to obtain the breakup of millimetric bubbles and keep them within the field of view long enough so that the size distribution can be recorded. In exchange for this large intensity, the turbulent flow field isn't homogeneous or isotropic, but we don't expect those features to affect our conclusions. 

% \begin{table}
%   \begin{center}
% \def~{\hphantom{0}}
%   \begin{tabular}{lccccccccc}
%       Surfactant &
%       Solvent &
%       Concentration (\% CMC) &
%       $\gamma_s$ (mN/m) &
%       $d_H$ (mm) &
%       $d_{32}$ (mm) &
%       $L_{int}$ (cm) &
%       $\eta$ ($\mu$m) &
%       $\varepsilon$ (m$^2$s$^{-3}$) &
%       $T_{int}$ (s) \\[3pt]

%       SDS & DI water              & 0--80   & 70--37 & 1.1--1.6 & 4.4--5.4 & 1.2--1.6 & 23--31 & 1.1--3.7 & 0.03--0.05 \\
%       SDS & artificial sea water  & 0--1.2  & 65--39 & 1.1--1.5 & 4.4--5.7 & 1.2--1.6 & 23--31 & 1.1--3.7 & 0.03--0.05 \\
%       Tween20 & DI water          & 0--100  & 72--41 & 1.2--1.7 & 5.7--6.1 & 1.2--1.6 & 23--31 & 1.1--3.7 & 0.03--0.05 \\
%       SDS & tap water             & 0--2    & 72--55 & 1.3--1.8 & 5.7--6.5 & 1.5--2.3 & 23--38 & 0.5--3.3 & 0.04--0.09 \\

%   \end{tabular}
%   \caption{Summary of surfactant and solvent conditions with corresponding fluid and turbulence parameters.}
%   \label{table:1}
%   \end{center}
% \end{table}

\section{Effect of surfactant on bubble production across scales}
\subsection{Representative break-up dynamics with and without surfactant}

\begin{figure}
    \centering
    \includegraphics[width=\linewidth]{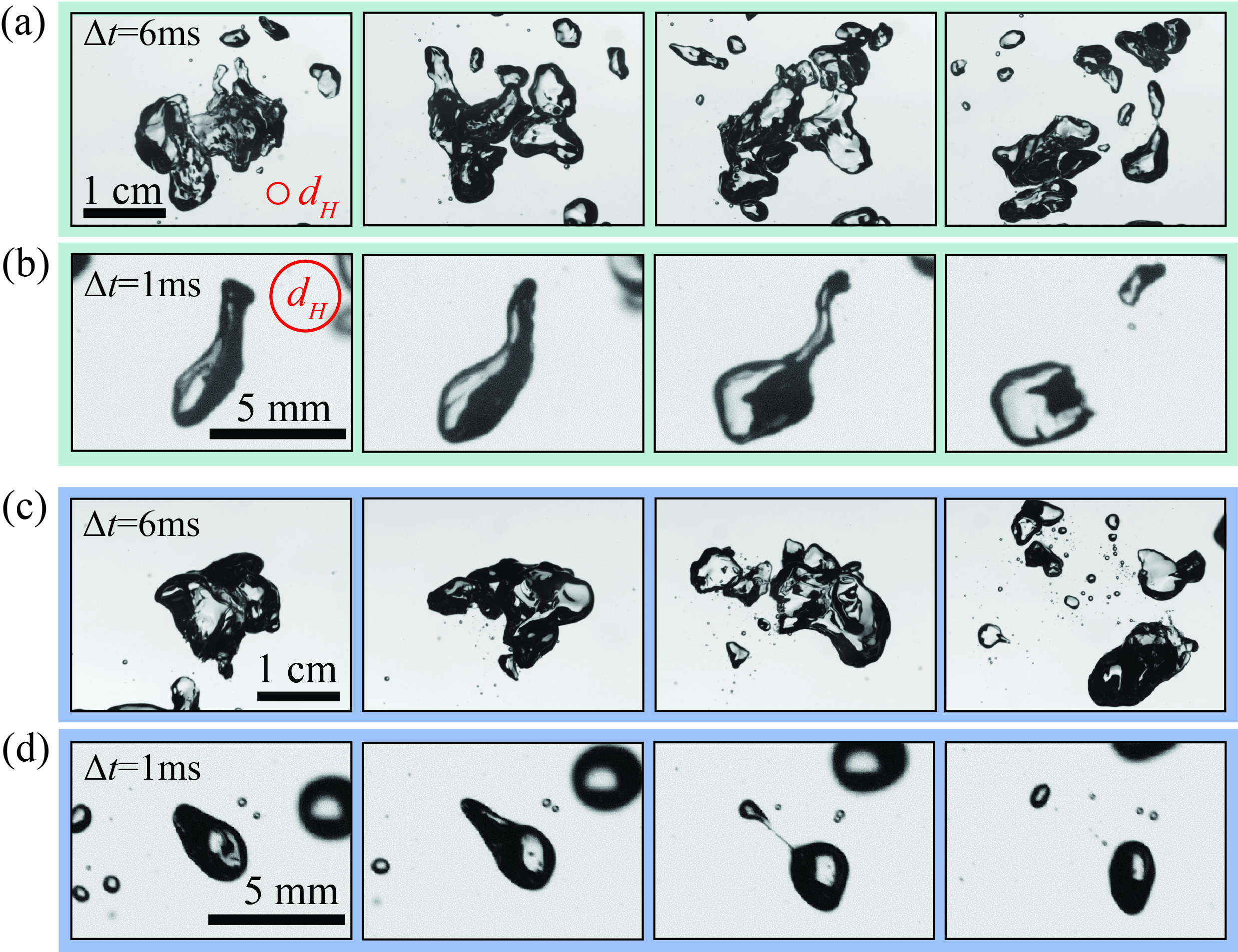}
    \caption{Representative image sequences illustrating bubble breakup with filament fragmentation and subsequent sub-Hinze bubble generation in presence and absence of surfactant. The initial parent bubble has a size on the order of the Hinze scale. Each row presents four time-resolved snapshots, with a temporal spacing of 1 ms (a, c) or 6 ms (b, d) between consecutive frames.
    (a) and (b): DI water without surfactant, showing the formation of no or a single sub-Hinze bubble.
    (c) and (d): 4000~\textmu mol/L of SDS, where several sub-Hinze bubbles are produced.}
    \label{fig:4}
\end{figure}

We qualitatively discuss examples of large bubble break-up in our experiments, leading to populations of sub-Hinze scale bubbles, in presence and absence of surfactant. Figure \ref{fig:4} presents representative examples of the fragmentation process with configurations showing filament formation or shedding behind large bubbles (without surfactants: (a) and (b) and with $4000~\text{\textmu mol}/\text{L}$ of SDS: (c) and (d)).

\citet{Riviere2021,Riviere2022,Ruth2022, qi2022fragmentation} discuss large bubble breakup in turbulence as happening through two successive stages. First, the parent bubble is deformed by a turbulent eddy with a characteristic length scale comparable to the bubble diameter, over a time-scale controlled by the eddy turn-over time ($T_{\text{eddy}} \sim 25 \text{ ms}$ in figure \ref{fig:4} (b,d). This interaction stretches the bubble into an unstable configuration consisting of two lobes connected by a thinning neck, as illustrated in figure \ref{fig:4} (b, d). Subsequently, the neck undergoes further thinning under the action of capillary forces, sometimes forming a filament, which ultimately breaks via a Rayleigh–Plateau like instability, over the capillary timescale ($T_{\text{cap}}\sim 0.1-1$ ms based on the filament thickness in figure \ref{fig:4} b, d), producing small capillary bubbles.

%The fragmentation is visibly influenced by the presence of surfactants. In pure water, the representative sequence (figure \ref{fig:4} b) is similar to the ones shown in \cite{Riviere2022,Ruth2022}, with one or two small bubbles formed following large bubble deformation. In surfactant-laden conditions, a similar configuration (figure \ref{fig:4} d) leads to the formation of a larger number of small child bubbles. 

The representations of the sub-Hinze bubble production in figure \ref{fig:4} (b) and (d) could suggest that surfactants influence this mode of production, by increasing the number of child bubbles formed by a single air filament. This configuration is however an idealization of the multiple concurrent fragmentations occurring with the large bubbles used in this study (The injection scale is $d_{\rm inj} \approx 40 d_H $) visible in figure \ref{fig:4} (a) and (c). In those case a single large bubble can be seen fragmenting into a large number of smaller bubbles, with many fragmentation events occurring between subsequent frames. One key difference is the resulting distribution of bubbles: in the clean case a few intermediate sized bubbles (around $d_H$) are formed as well as a similar number of sub-Hinze bubbles. In the contaminated case however, a similar number of intermediate sized bubbles are formed, but the number of sub-Hinze bubble is drastically larger. 

%This can be confirmed by considering the coalescence rate decided by turbulence motion: R(a)=2*pi*(2a)^3*N(a)/t_k, where R(a) is the rate for collision of a given particle a, N(a) is the particle density and t_k is the Kolmogorov timescale. In our study, the rate of collision for sub-Hinze bubbles is smaller than 1, which verifies that the coalescence effect is minimal.
}
%We comment, however, that our time resolution and the use of a single camera is not sufficient to access detailed dynamics in a reliable way. Experiments in a more controlled setting, where the bubble always breaks at the same spot (using pinch off setups, or single vortex breakup setups \cite{Ruth_2019,ni2024deformation} and higher temporal resolution with multiple cameras would be more adequate. Our setup was design to consider statistics of bubble sizes In the next section, we will quantitatively analyze the effect of surfactants on the bubble size distribution resulting from the fragmentation of large air cavities.

Different mechanisms may account for the enhanced production of small bubbles in the surfactant-laden case. Surfactants may increase the number of small child bubbles produced during the deformation and breakup of large bubbles in a tearing or shedding like process or during the rupture of elongated individual ligaments. Distinguishing between these mechanisms would require dynamically resolved experiments or simulations focusing on these different scenarios.
We comment, however, that our time resolution and the use of a single camera is not sufficient to access detailed dynamics in a reliable way. Our study focuses on the statistical analysis of the influence of surfactant on the fragmentation of a relatively large air cavity in turbulence. Detailed observation of the break-up dynamics would require zoomed in views at higher temporal resolution, with a dedicated experimental setup with multiple cameras such as some past studies (without surfactant) on pinch-off in turbulence \citep{Ruth_2019} or break-up by a single vortex \citep{qi2022fragmentation}. 

%\textcolor{orange}{In addition, this work focuses on the transient stage during which a large bubble undergoes successive fragmentation into smaller bubbles. The duration of the experiment may be shorter than the adsorption timescale of Tween 20. However, the timescale investigated here is comparable to that relevant to bubble entrainment and fragmentation beneath breaking waves.} 

%\hl{The ligament fragmentation} process is visibly influenced by the presence of surfactants. In pure DI water (figure \ref{fig:4} a), the ligament remains relatively thick, with a diameter comparable to that of the parent bubble, and typically produces at most one capillary bubble upon breakup. In contrast, in surfactant-laden conditions (figure \ref{fig:4} b), the ligament becomes significantly thinner and its subsequent pinch-off results in the formation of a cascade of capillary bubbles. In the presence of surfactants, the resulting capillary bubbles are substantially smaller than those formed in pure DI water. In the next section, we'll quantitatively analyze the effect of surfactants on the bubble size distribution resulting from the fragmentation of large air cavities.

\subsection{Bubble size distribution}

\begin{figure}
    \centering
    \includegraphics[width=\linewidth]{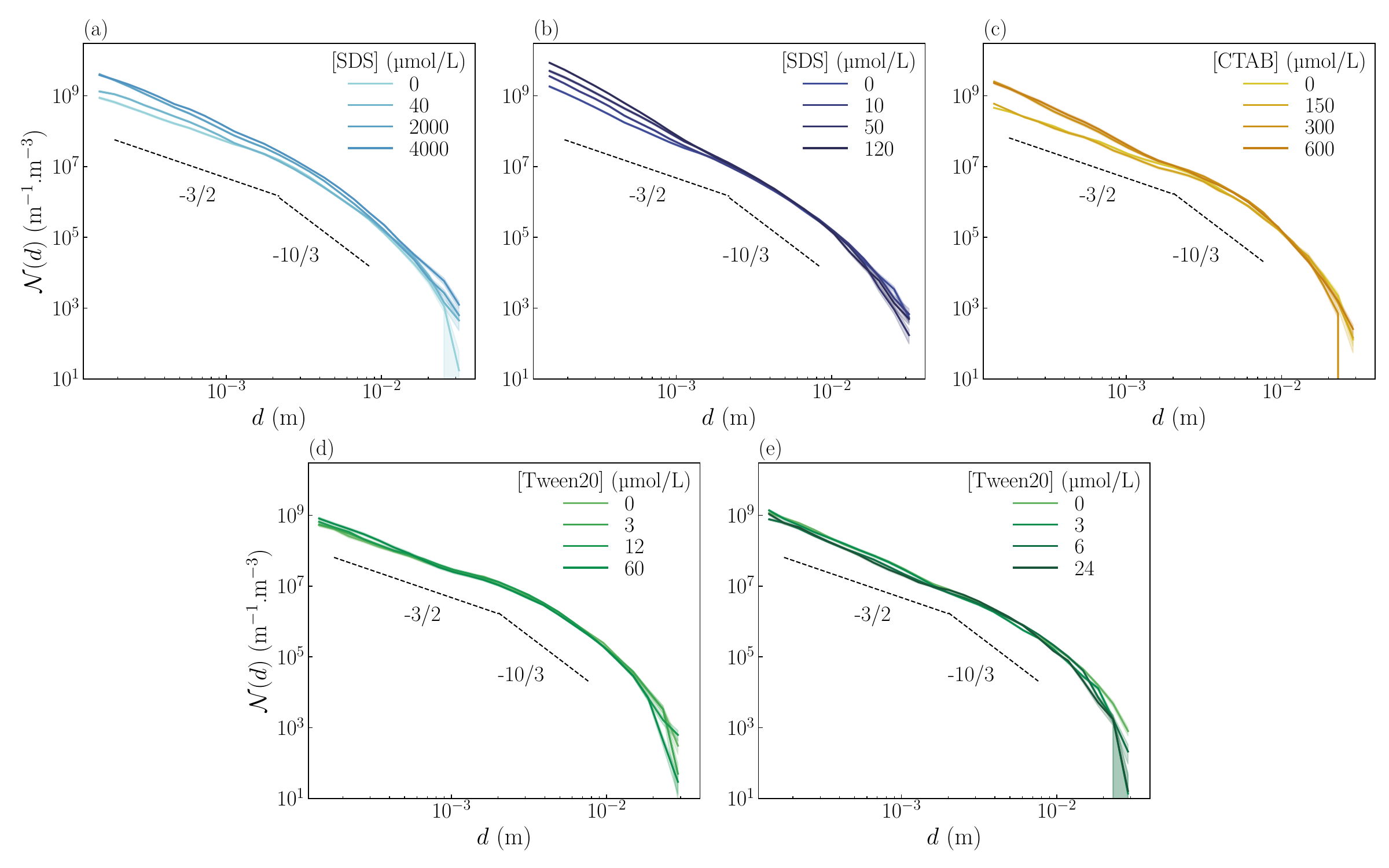}
    \caption{
    Comparison between the bubble size distributions of different solution composition at varying surfactant concentration. Bubble size distribution for (a) SDS, (b) SDS and sea salt, (c) CTAB, (d) Tween20, (e) Tween 20 and sea salt. Colors indicate the surfactant concentration. Dashed lines indicate the $d^{-3/2}$ and $d^{10/3}$ power laws. The shaded regions represent uncertainties in the distributions, estimated using a Monte Carlo approach, while the solid lines indicate the mean values in the distributions.}
    \label{fig:5}
\end{figure}

Figure~\ref{fig:5} presents the bubble size distribution $\mathcal{N}(d)$ for the tested solutions: SDS (a), SDS with sea salt (b), CTAB (c), Tween~20 (d) and Tween~20 with sea salt (e) at increasing surfactant concentrations. 
The dashed lines are shown as references for the $-10/3$ scaling in the super-Hinze regime and the $-3/2$ scaling in the sub-Hinze regime. The observed transition between these two regimes in the bubble size distribution occurs at $d \approx 2.5$ mm, which is close to the calculated Hinze scale, and will be used as effective Hinze scale $d_{H,\mathrm{eff}}$. 

%\textcolor{red}{Although the Hinze scale $d_H$, calculated from (\ref{eq:dh}) and reported in the fifth column of table~\ref{table:1}, is expected to vary with the static surface tension $\gamma_s$ and thus shift the bubble size distribution, this variation is negligible in practice. The effective Hinze scale remains approximately constant across all cases, with $d_{H,\mathrm{eff}} \approx 2.5~\text{mm}$.}

%A Monte Carlo approach was employed, in which data were randomly sampled from both the wide-angle and K2 datasets to compute $\mathcal{N}{\mathrm{wa}}(d)$ and $\mathcal{N}{\mathrm{k2}}(d)$ using Eq.~\ref{eq:1}, repeated over 100 realizations. The mean and uncertainty of $\mathcal{N}(d)$ were then obtained from the ensemble of realizations. To generate a single, continuous distribution across the entire size range, $\mathcal{N}{\mathrm{wa}}(d)$ and $\mathcal{N}{\mathrm{k2}}(d)$ were combined using a smooth transition function:
%\begin{equation}
%\mathcal{N}(d)
%=\tfrac{1}{2}\left[1+\tanh\left(\frac{d-d_c}{w}\right)\right]\mathcal{N}{\mathrm{wa}}(d)
%+\Bigg[1-\tfrac{1}{2}\left[1+\tanh\left(\frac{d-d_c}{w}\right)\right]\Bigg]\mathcal{N}{\mathrm{k2}}(d),
%\end{equation}
%where $d_c = 1~\mathrm{mm}$ and $w = 0.5~\mathrm{mm}$ are parameters chosen to ensure a smooth transition between the two datasets.

For a given set of experiments, the distribution in the super-Hinze scale range ($d > d_{H,\mathrm{eff}}$) does not change when adding surfactants, and is very similar across the different solutions tested. As the concentration of SDS is increased, with or without sea salt (Figure~\ref{fig:5} a and b), the total number of bubbles in the sub-Hinze range increases while that in the super-Hinze regime remains unchanged. The increase in the number of bubbles in the sub-Hinze range is associated with an increase in the slope of the distribution. 
In contrast, when Tween20 is used (Figure~\ref{fig:5} d and e), the bubble size distribution remains unchanged both in the super-Hinze and sub-Hinze ranges. This conclusion is not altered by the presence of sea salt. Using CTAB leads to an intermediate behaviour between that of SDS and Tween~20, where the number of sub-Hinze bubbles increases significantly once the concentration is large enough. Details of the solution (salt, surfactant) only affect sub-Hinze scale bubbles, and the effect is strongly dependent on the type of surfactant used, while the super-Hinze size distribution remains unchanged.

For all conditions, the power law distribution in the sub-Hinze regime at the lowest contamination is compatible with the $d^{-3/2}$ distribution due to capillary break-up. For the cases that show a increase in the number of sub-Hinze bubble count, this increase is linked to a steeper distribution as surfactant concentration is increased (i.e. a more negative power law exponent). This modification of the shape of the distribution, together with the fact that we do not observe any changes in the effective Hinze scale when adding surfactant, rules out the possibility that the trends observed in this study are merely a result of increasing the Weber number through a change of the static surface tension.

%The inset in Figure~\ref{fig:3} (c) shows the variation of the fitted slopes $\alpha$. %(uncertainty on the fitting slope is obtained by a Monte Carlo approach randomly varying the starting and endpoint of the size interval in a reasonable range). 
%a starting point within $d \in [150~\mu\mathrm{m}, 400~\mu\mathrm{m}]$ and an endpoint within $d \in [2~\mathrm{mm}, 3.5~\mathrm{mm}]$ in Figure~\ref{fig:3}, and performing 1000 repetitions of the fitting to obtain the mean and standard deviation of $\alpha$. 
%For SDS in DI water (blue), artificial seawater (darker blue), and CTAB in DI water (yellow), the fitted slopes $\alpha$ become more negative as contamination increases whereas for Tween20 in DI water (green), the slope remains nearly constant. These observations are consistent with visual observation of the distributions which do not show changes for Tween20 across surfactant concentrations while more bubbles are observed in the sub-Hinze regime for SDS solutions. Note that this rules out the possibility that the trends observed in this study are merely a result of increasing the Weber number through a change of the static surface tension. 

%Indeed, if this was the case, the whole distribution would be shifted upwards, with the same exponent -3/2 in the sub-Hinze regime. Instead, we expect Marangoni effects during the capillary instability of the air filaments to play a key role in setting the distribution.

\subsection{Counts of Super-Hinze and Sub-Hinze bubbles}
We now quantify the increase of sub-Hinze bubbles ($N_{d<d_H}$) for the various solutions used. %$\mathrm{N}_{d>d_H}$ and below 

\begin{figure}
    \centering
    \includegraphics[width=0.9\linewidth]{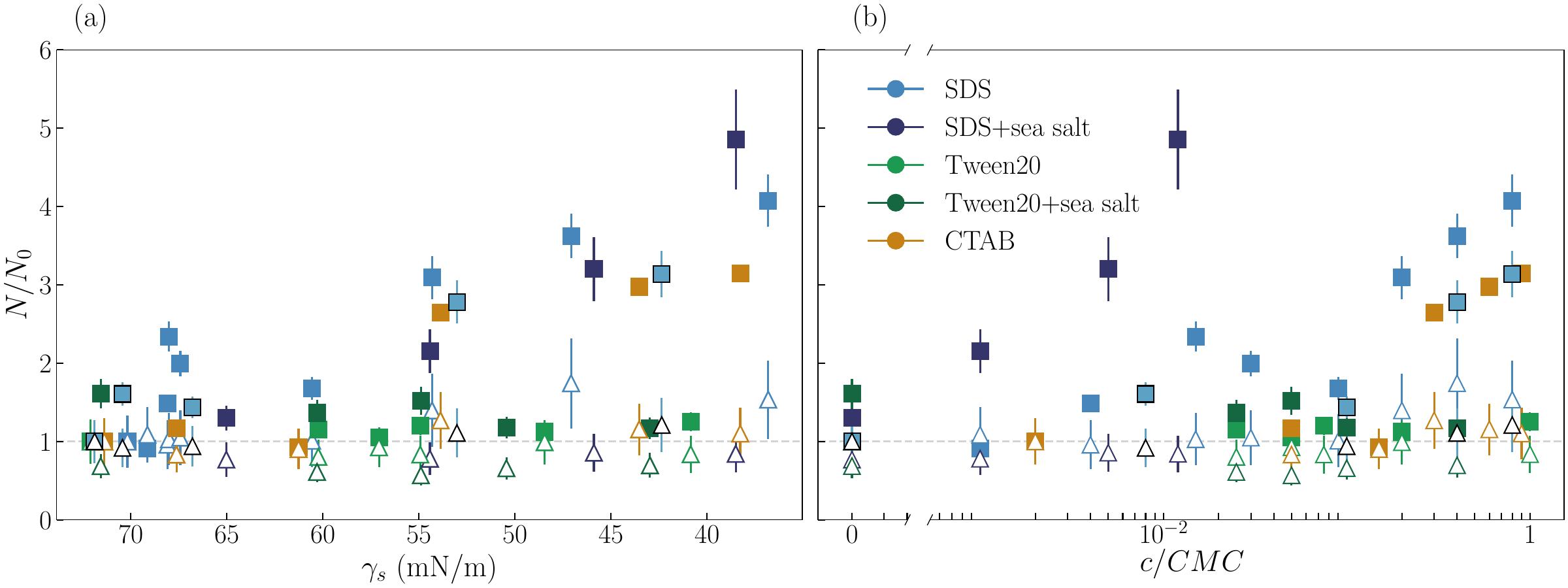}
    \caption{(a) Variation in the counts of super-Hinze (open triangles) and sub-Hinze (solid squares) bubbles with surface tension. The solid black edge denotes the second realization of SDS experiment. (b) Same data as a function of the bulk concentration of surfactant, represented as a fraction of the CMC.}
    \label{fig:6}
\end{figure}

Figure~\ref{fig:6}(a) shows the counts of super-Hinze bubbles $N_{d>d_H}$ (open triangles) and sub-Hinze bubbles $N_{d<d_H}$ (full squares) as contamination increases (represented by the static surface tension, note the inverted x-axis) for the different types of contamination (colors). We define the reference bubble count, $N_0$, as the bubble count measured in pure water without surfactant or salt. With this definition, the normalized bubble count, $N/N_0$, collapses to unity at zero surfactant concentration (corresponding to the maximum static surface tension, $\gamma_s$) for SDS, Tween~20 and CTAB. For surfactant solutions including sea salt, however, the initial normalized counts are slightly greater than unity, possibly due to trace amounts of surfactants present in the sea salt that we use. Figure~\ref{fig:6}(b) shows the same data as a function of the bulk concentration of surfactant, represented as a fraction of the CMC. % The second realization of SDS in DI water is indicated by light blue squares outlined in black.

As contamination increases with the addition of surfactant (surface tension is reduced), the super-Hinze bubble count remains unchanged in all cases (triangles). For sub-Hinze bubbles, all solutions with a fast surfactant (for instance SDS, blue squares) show an increase in counts: for very low surface tensions, the counts of sub-Hinze bubbles are increased a factor 5. In contrast, for cases where surface tension is modified by adding a slow surfactant (Tween~20, green squares), the sub-Hinze bubble counts remain unchanged. The addition of salt is associated with a similar trend: the number of sub-Hinze bubbles increases with increasing SDS concentration (dark blue squares), while remaining essentially unchanged with increasing Tween~20 concentration (dark green squares). The intermediate surfactant (CTAB, yellow squares) exhibits a transitional behaviour: it behaves like a slow surfactant at higher static surface tension $\gamma_s$ (low concentration), and increasingly resembles a fast surfactant as $\gamma_s$ further decreases (high concentration). In particular, as the CTAB concentration is increased, the sub-Hinze bubble count remains unchanged for $\gamma_s \gtrsim 60~\text{mN}\text{m}^{-1}$, before increasing as $\gamma_s$ is further reduced. The concentration of surfactants required to obtain a similar increase in counts can be vastly different depending the the surfactant used or on whether sea salt is present (figure \ref{fig:6} b), suggesting a reinforcement of the effect of SDS (anionic) upon the addition of ions. Another way to view the same effect is is the effective reduction of the CMC of charged surfactants in the presence of ions: indeed the CMC of SDS is reduced by almost two orders of magnitude when adding NaCl at seawater concentration \citep{schick_effect_1964,srinivasan_effect_2003}, comparable to the differences seen in Fig. \ref{fig:6} (b). This effect is also relatively well be captured by plotting the counts as a function of the static surface tension (figure \ref{fig:6} a).
%Remarkably, all the contaminated solutions with SDS are very close one to each other suggesting that the presence of salts or tap water has little effect on the break-up process, with the presence of a rapid adsorption surfactant such as SDS dominating the influence on the break-up process. 

As stated above, the various cases with Tween 20 show no differences in counts when varying the concentration despite a reduction in surface tension comparable to the reduction when using SDS, highlighting that the increase in the number of sub-Hinze bubbles and the change in the power law observed here is not simply due to a reduction in surface tension and associated increase in Weber number. Tween 20 is non ionic and interactions between the surfactant and sea salt are therefore very limited.

\begin{figure}
    \centering
    \includegraphics[width=0.8\linewidth]{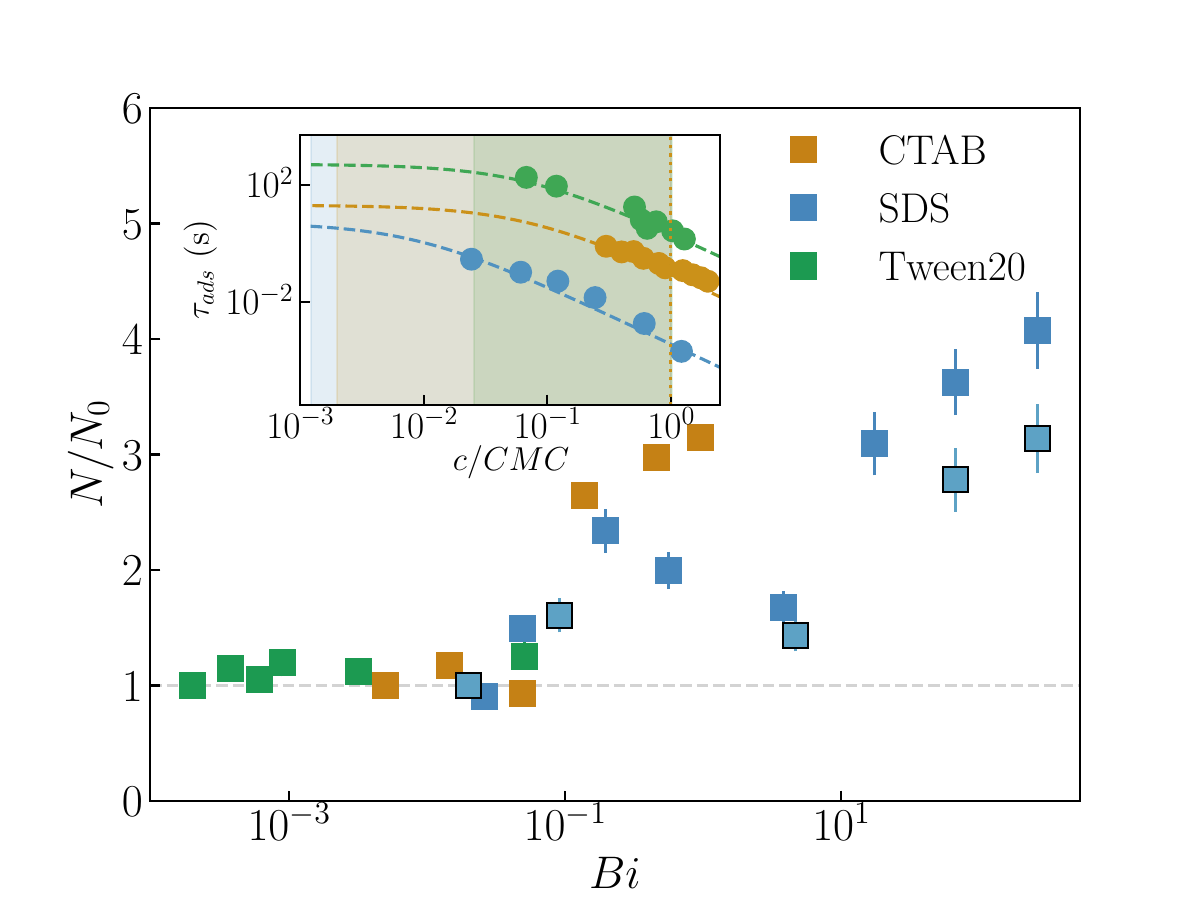}
    \caption{Variation in the number of sub-Hinze bubbles as a function of the Biot number $Bi$ for SDS (blue), CTAB (yellow), and Tween 20 (green). The second SDS realization is denoted with a black edge. Inset: Adsorption time $\tau_{\rm ads}$ as a function of bulk concentration of surfactant in solution. Data points are from the literature, shaded areas represent the ranges used in this study and dashed lines are fits used to extend estimated adsorption times to lower concentrations}. Yellow for CTAB \citep{Qazi2020}, blue for SDS \citep{Fainerman2010} and green for Tween 80 \citep{Qazi2020}.
    \label{fig:7}
\end{figure}

As hinted in the introduction and now demonstrated experimentally, the influence of surfactant on the break-up process is tightly linked to their adsorption time. Surfactants will only affect the breakup process if a significant amount can adsorb to the interface of a bubble between two successive fragmentation events. Indeed, during the stretching preceding fragmentation (through tearing or through the formation of a ligament), a large amount of surface area is created and therefore the surfactant surface concentration on a newly formed bubble is less than the value at equilibrium with the bulk (see sketch in figure \ref{fig:1} b). If the timescale for adsorption of surfactants to the interface $\tau_{\rm ads}$ is much longer than the typical time between fragmentation events, then after a few breakups, the bubbles will essentially behave as clean bubbles by successive dilutions. We consider here a large scale turbulent time: the eddy turn over time at the integral length scale $T_{\rm eddy}(L_{\rm int})$ (equal to the integral time scale $T_{\text{int}}$) as the representative time for the turbulence break-up sequence to be compared to the adsorption process. To quantify the competition between turbulent transport and adsorption kinetics, we introduce a Biot number \citep{tasoglu2008effect,pico2024surfactant}, comparing the turbulence timescale to the adsorption timescale:
\begin{equation}
    Bi = \frac{T_{\rm eddy}(L_{\rm int})}{\tau_{\rm ads}}. %= \frac{d^{2/3} T_{\rm int}}{L_{\rm int}^{2/3} \tau_{\rm ads}}.
\end{equation}
%In the following, we use $d_{H, {\rm eff}}\simeq2.5$~mm as a typical length scale to compute the Biot number (a different value would simply shift the estimated $Bi$ without changing our conclusions).
We therefore need to estimate the adsorption timescale for the various contamination cases considered. The inset of figure \ref{fig:7} shows data acquired using a maximum bubble pressure tensiometer for SDS (extracted from \cite{Fainerman2010}), Tween 80 (a surfactant similar to Tween 20 but with more oxyethylene groups, extracted from \cite{Qazi2020}) and CTAB (also extracted from \cite{Qazi2020}). The adsorption timescale differs markedly across surfactants, with Tween~80 ($\sim 10~\text{s}$) being much slower than SDS ($\sim 10~\text{ms}$), while CTAB lies in between ($\sim 1~\text{s}$). In all cases, the adsorption timescales become much faster as concentration is increased, with a variation close to $\tau_{\rm ads} \propto c^{-2}$. Note however that the datasets cover concentrations that are larger than the ones we use (highlighted with the shaded areas). We therefore need to extend $\tau_{\rm ads}$ to lower concentrations via modeling.

We follow the approach detailed by \citet{Ferri2000}: under the assumption that the adsorption is diffusion limited, the adsorption timescale can be expressed as $\tau_{\rm ads} \sim h^2 / D$, where $D$ is the diffusivity of the surfactant and $h$ is the depletion depth, the thickness of the layer depleted of surfactants below the interface. This depth is then given by $h = \Gamma / c$, with $c$ the bulk concentration and $\Gamma$ the surface concentration at equilibrium, which is not directly accessible but given by the isotherm (or surfactant equation of state) that the surfactant follows. We make the hypothesis that both surfactants follow a Langmuir isotherm, in which case $\Gamma(c) = \Gamma_\infty K c / (1 + K c)$, with $\Gamma_\infty$ and $K$ two constants. Bringing these results together, we obtain finally that the adsorption timescale should follow $\tau_{\rm ads} = t_0 / (1 + c / c_0)^2$, with $t_0$ and $c_0$ two fitting parameters. 
This functional form fits well the available literature data, in particular the $c^{-2}$ scaling observed in all datasets (dashed lines). Note however that the approach that we use here is only done to extract a typical adsorption time and has several limitations: first, we have no proof that the adsorption is indeed diffusion limited and that kinetics do not start to play a role for some of the concentrations (although if diffusion is the limiting factor, one expects the $c^{-2}$ scaling observed in the data \citep{ward1946time}). Second, both datasets were recorded in quiescent conditions, whereas in our case turbulence constantly mixes the fluid and brings surfactant close to the interface. As a consequence, diffusion may be significantly faster than expected (or the depletion depth thinner) and adsorption kinetics may play a role \citep{chang1995adsorption}. 

We can now plot the normalized count of sub-Hinze bubbles as a function of the Biot number, computed using the fitted adsorption timescale in the inset of figure \ref{fig:7}. As expected, we now very clearly see two different regimes: at low $Bi$, the surfactant is too slow for a significant amount to be adsorbed between breakup events and as a consequence the dynamics is identical to the clean case ($N/N_0\approx1$). At higher $Bi$ however, the number of sub-Hinze bubbles progressively increases as a significant amount of surfactants can adsorb to bubbles before they fragment through a capillary instability, during which important surface tension gradients can then form and influence the resulting size distribution. We note that using another turbulent time would lead to a similar plot, simply shifting the transitional Biot number. With the present choice of the slowest turbulent time, the transition is for $Bi$ between 0.1 and 1.

\section{Conclusion}

Surface active compounds can have a very large impact on bubble breakup in turbulence: they drastically increase the production of sub-Hinze scale bubbles (by up to a factor five) while leaving the number of super-Hinze bubbles unchanged.  The size distribution of bubbles is modified with the sub-Hinze range still well described by a power law, but with an exponent increasingly more negative than -3/2 as the contamination of the solution is increased. This steepening may have important consequences for processes sensitive to the number of small bubbles, e.g air-sea gas exchange or sea salt aerosol production.

However not all surfactants show the effects described above. As demonstrated experimentally, Tween 20, despite reducing the surface tension similarly to SDS, has strictly no effect on the size distribution of bubbles. CTAB exhibits transitional behaviour, with an increase in sub-Hinze bubble production observed only for large concentrations (or when the static surface tension is sufficiently low). This striking difference is due to the slow adsorbtion of Tween 20 compared with a typical time of deformation of the large bubbles in the turbulent flow, quantified with the Biot number. 
The addition of ions in the solution reinforces the effects of SDS and a concentration a hundred times smaller can have the same effect on the breakup process as SDS only when artificial sea salt is added, for comparable static surface tension.
Tween~20 remains ineffective in modifying sub-Hinze bubble production, even in the presence of ions. Our results therefore demonstrate that static surface tension is not the only parameter to discuss the effects of surfactants on fragmentation. In other interfacial fluid dynamics problems, we highlighted the importance of Marangoni effects and gradients in surface tension that can be characterized by Langmuir trough measurements of isotherms under compression \citep{Erinin2023, eshimaauregan2025}. Here the measured isotherms for the Tween 20 or SDS solutions are fairly similar (see App. \ref{sec:langmuir}) and a metric based on the Marangoni number alone would not explain the differences in the sub-Hinze bubble production.
The common denominator between the problems cited above is that the effects of surfactants could be treated by considering an initially contaminated interface with insoluble surfactants. Because of the solubility of the surfactants considered here, and of the successive fragmentations that bubbles will undergo during the experiment, this approach is not sufficient.
It therefore remains an open problem to find the right sets of parameter to fully describe a solution and predict the final size distribution of bubbles. These parameters should take into account the surfactants adsorption time, possible ionic interactions, and the ability for a given surfactant to generate strong Marangoni flows, as well as clarify the detailed mechanism responsible for the increase in small bubbles due to surfactants.

\textbf{Declaration of Interests}~ {The authors report no conflict of interest.}

\textbf{Acknowledgments}~{We thank Emmy Roy who conducted preliminary experiments and Jun Eshima for discussion. We thank the reviewers for their helpful comments.}

\textbf{Funding}~{This work was supported by NSF grant 2242512 to L.D.; the Princeton Catalysis Initiative and the Moore Foundation.}

\appendix
\section{Langmuir trough measurement}\label{sec:langmuir}
\begin{figure}
    \centering
    \includegraphics[width=0.9\linewidth]{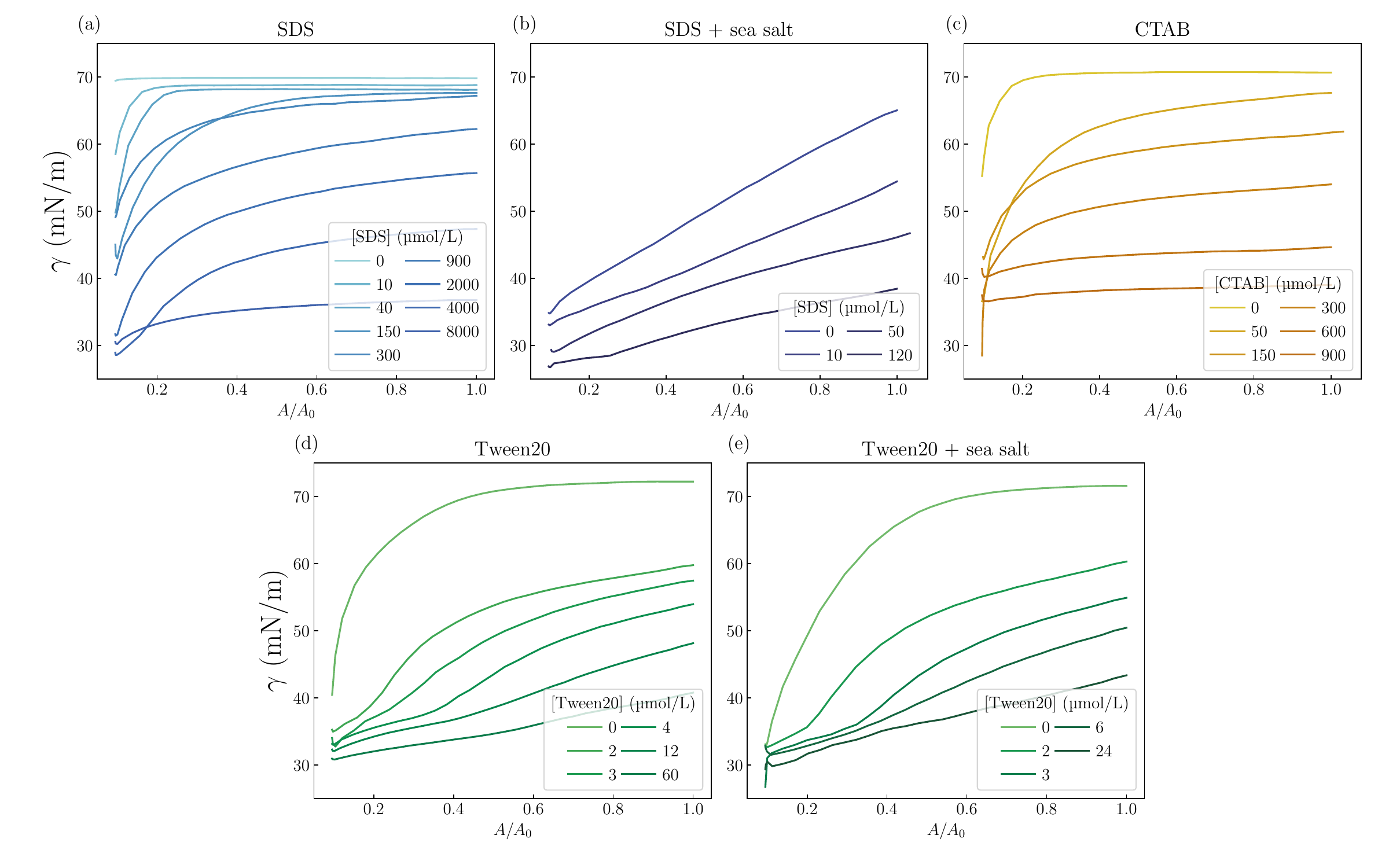}
    \caption{Surface tension isotherms of the various solutions used in this study. (a) SDS, (b) SDS and sea salt, (c) CTAB, (d) Tween 20, and (e) Tween 20 and sea salt.}
    \label{fig:langmuir}
\end{figure}

For each experimental case, the solution’s surface tension isotherm was measured at the beginning and end of the experiment using a Langmuir trough (KSV NIMA 1003). Specifically, we record the properties of the surface by setting aside a small amount of solution and placing it in the trough, and the surface tension is measured by a force balance with a platinum Wilhelmy plate. Two Teflon barriers compress the surface at a rate of 270 mm/min allowing us to obtain the surface tension $\gamma_s$ as a function of the trough area. The resulting data are plotted in figure \ref{fig:langmuir} as a function of the area, normalized by the area before compression $A_0$.

The range of SDS concentrations tested (figure~\ref{fig:langmuir}a) spans the full range of interfacial properties, from very clean conditions (flat isotherm at $72~~\text{mN}$ $\text{m}^{-1}$) to a nearly saturated interface (approximately flat isotherm at $\sim 40~\text{mN}$ $\text{m}^{-1}$, corresponding to $80\%$ of the CMC). The SDS isotherms with sea salt added (figure~\ref{fig:langmuir}b) also cover a broad range of surface tension values, but over a much narrower concentration range (up to $1.2\%$ of the CMC of SDS). The concentrations used with CTAB range up to the CMC and yield static surface tension values range from 72 to less than 40 mN/m.
For Tween~20, we use concentrations up to the CMC, yielding a similar range of static surface tension, $\gamma_s$, to that of the SDS cases (figure~\ref{fig:langmuir} d). In contrast to SDS, the concentrations required to obtain a low surface tension are similar with and without sea salt. 
%It should be noted that, due to surface aging, the surface tension of Tween~20 solutions decreases with time following the creation of a new interface when pouring into langmuir trough. To minimize this effect, isotherms were measured as rapidly as possible after introducing the solution into the trough. This aging effect attenuates at higher Tween~20 concentrations. However, surface aging does not affect the bubble breakup results: experiments repeated after a sufficiently long waiting time (5 min) yield identical outcomes. The isotherms of CTAB in DI water are shown in figure \ref{fig:langmuir} (e).

\section{Time evolution of bubble statistics}
\begin{figure}
    \centering
    \includegraphics[width=0.9\linewidth]{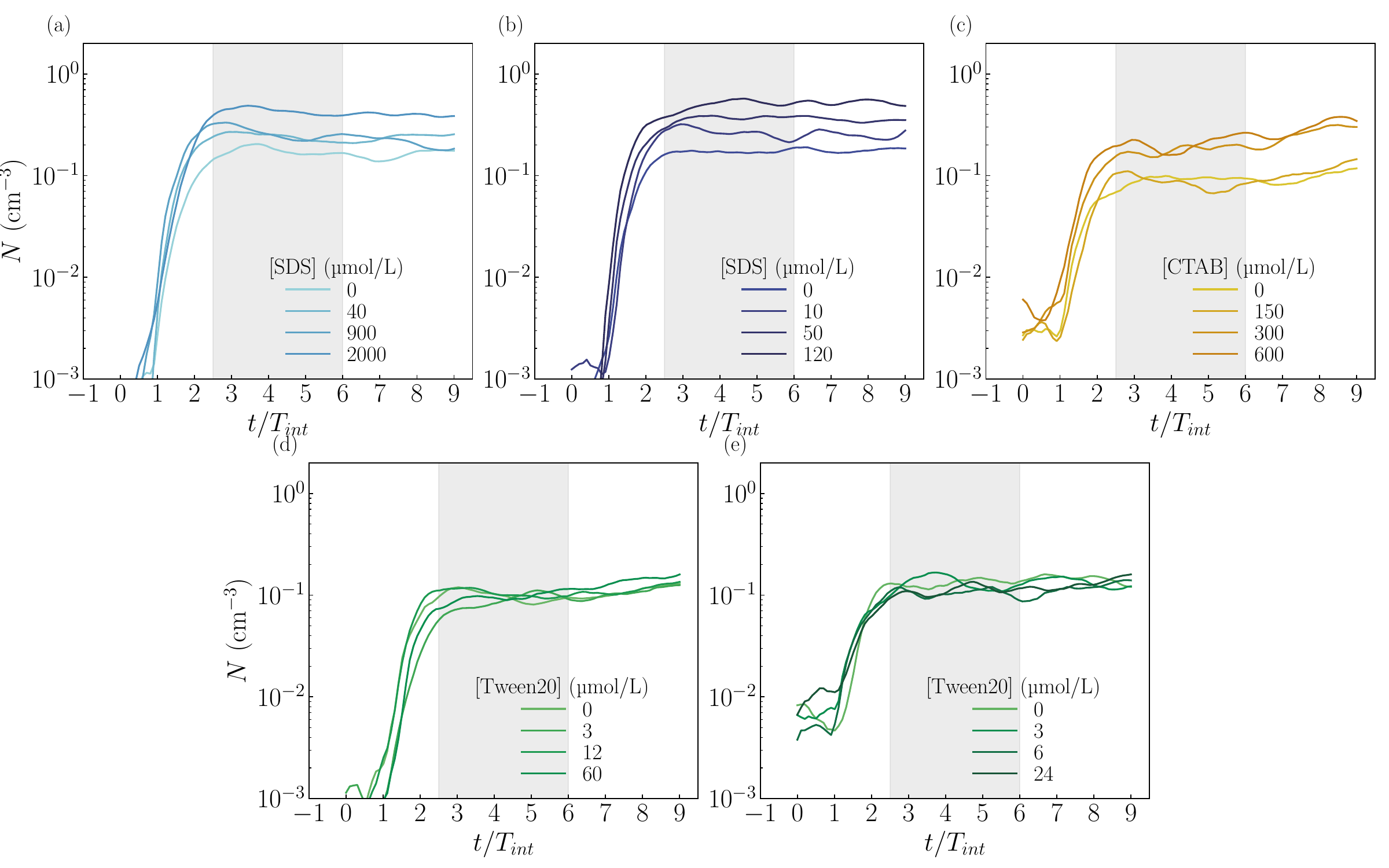}
    \caption{The evolution of the bubble number of (a) SDS, (b) SDS and sea salt, (c) CTAB, (d) Tween 20, and (e) Tween 20 and sea salt. The gray area represents the range used to compute the distributions in the rest of the study.}
    \label{fig:nevolve}
\end{figure}
Figure \ref{fig:nevolve} shows the temporal evolution of bubble counts $N = \int \mathcal{N}(d)\mathrm{d}d$ over time for all the cases studied. Across all experimental conditions, $N$ exhibits a sharp increase immediately after the cup is flipped ($t=0$), followed by a plateau until the end of the recording. The presence of this plateau shows that number of bubbles entering and leaving the field of view balances out on average. The increased number of sub-Hinze bubbles described in the main text is visible on these graphs.

%We note that the experiment is transient by design as a large cavity is released and breaks due to turbulence. In our experiment, we observe bubbles generated over a relatively short window after the release ($\leq$ 1 s, about 10 $T_{\rm int}$). During that time, bubbles rise to the surface and break and we observe an initial increase in the observed bubbles which plateaus for some time, and would then decrease at much longer times if we were continuing to record as all bubbles get out of the experimental field of view. 

%Even though the observation time is shorter than the adsorption time of Tween 20, the relevant comparison is not the time since the release but the time in between the break-ups in the sequence. As new area is constantly being created through successive breakups, the relevant timescale is the time in between break-ups.

\begin{figure}
    \centering
    \includegraphics[width=0.9\linewidth]{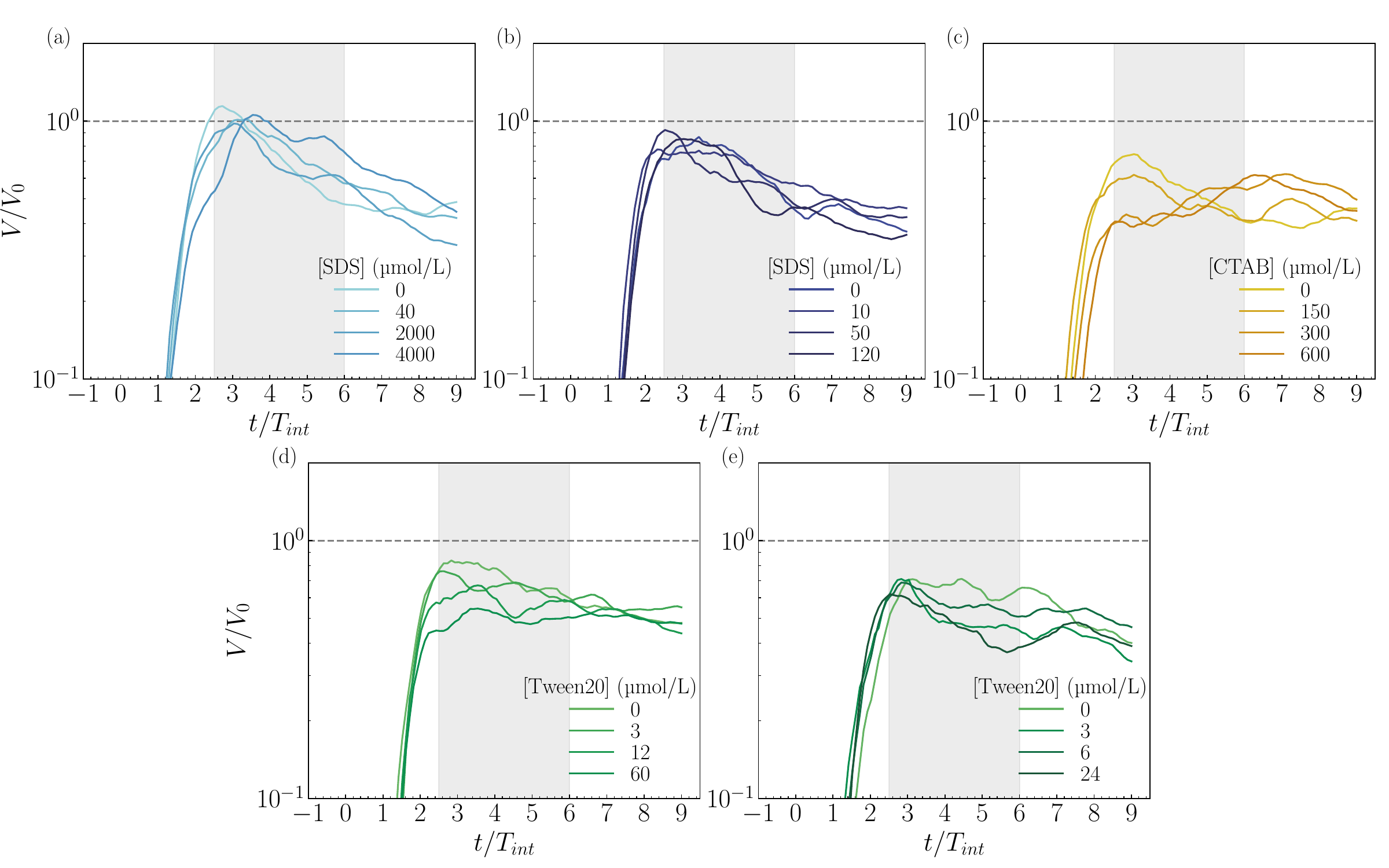}
    \caption{The evolution of the cumulative bubble volume of (a) SDS, (b) SDS and sea salt, (c) CTAB, (d) Tween 20, and (e) Tween 20 and sea salt. The gray area represents the range used to compute the distributions in the rest of the study. The horizontal dash line represents $V=V_0$.}
    \label{fig:vol}
\end{figure}

Figure \ref{fig:vol} shows the temporal evolution of the cumulative bubble volume $V = \frac{1}{6}\pi\int d^3\mathcal{N}(d)\mathrm{d}d$, at every moment for all cases. $V_0$ is the volume of air injected into the cup before the experiment. After the cup is turned around, nearly all the volume is detected. After that the total volume slowly decreases during the experiment due to the larger bubbles getting out of the field view due to buoyancy. 

In the main text, we average data over the interval $2.5 T_{\mathrm{int}} \le t \le 6 T_{\mathrm{int}}$, this range was chosen as a balance that allows us to obtain sufficiently converged statistics while retaining most of the large scale bubbles. Note that we also verified that the shape of the bubble size distribution does not vary significantly over that window. 

\end{document}